\documentclass[
aps,prd,showpacs,showkeys,amsmath,amssymb,
eqsecnum,preprintnumbers
]{revtex4}

\usepackage{dcolumn}
\usepackage{bm}
\usepackage{graphicx}

\begin{document}

\title{Surface holonomy and gauge 2-group
\footnote{Submitted upon invitation to {\it International Journal of
Geometric Methods in Modern Physics.}}}

\author{Amitabha Lahiri}


\email{amitabha@boson.bose.res.in}
\affiliation{S. N. Bose National Centre for Basic Sciences, \\
Block JD, Sector III, Salt Lake, Calcutta 700 098, INDIA\\}
\date{\today}

\preprint{{hep-th/0402227}}


\begin{abstract}
Just as point objects are parallel transported along curves, giving
holonomies, string-like objects are parallel transported along
surfaces, giving surface holonomies. Composition of these surfaces
correspond to products in a category theoretic generalization of the
gauge group, called a 2-group. I consider two different ways of
constructing surface holonomies, one by using a pair of one and two
form connections, and another by using a pair of one-form
connections. Both procedures result in the structure of a 2-group.
\end{abstract}


\maketitle

\section{Introduction}\label{intro}

The notion of parallel transport plays an important role in
physics.  Elementary particles and the fields whose point-like
excitations describe them, may be thought of as carrying
representations of Lie algebras. To describe the dynamics, vector
fields at nearby points must be compared, and parallel transport
provides a self-consistent way of doing so. When the particles are
charged, parallel transport requires the existence of a connection
on a fiber bundle constructed on space-time. This connection is
called a gauge field, and all interactions of particles can be
described by these gauge fields. The dynamics of gauge fields are
described by nonlinear (Yang-Mills) theories, and it is expected
that in the strong coupling regime there are extended objects, such
as strings, or flux tubes in the theory. So it is worthwhile to
consider parallel transport of string-like objects.

There is an immediate obstacle to parallel transporting `charged'
flux tubes. For particles, parallel transport involves carrying a
vector (in some Lie algebra) along a specified curve in
space-time. Consider an infinitesimal curve of length $\epsilon$
and tangent $\tau^\mu\,.$ Then parallel transport is a group
action, where the vector is acted upon by a group element of the
form $g(\epsilon, \tau, A) \sim 1 + \epsilon \tau^\mu A_\mu\,,$
where $A_\mu$ is the connection. Now, curves can be joined end to
end, infinitesimal curves can be joined to produce a finite
curve. Joining curves is the same as composition of the
corresponding group elements, and thus it becomes possible to
define the parallel transport of a vector along any finite curve in
terms of group elements. So given a starting point and a
connection, a curve can be uniquely identified with a group element
by taking the path ordered exponential of the gauge field $A$ along
that curve. This identification is called the holonomy of the
curve.

This procedure cannot be generalized to extended objects except
trivially. That is, suppose I could associate an element of some
group to the infinitesimal parallel transport of a string, a sort
of `surface holonomy'. For simplicity, let me take the string to be
infinitesimal as well. Now let me try to construct the parallel
transport of a finite string between two finitely separated
configurations, say with the same end points. I should break up a
surface bounded by these two configurations into infinitesimal
surface elements, bounded by infinitesimal portions of
`intermediate' configurations of the string.  Since I know the
surface holonomy of each little area element, can I compose them to
get the holonomy for the entire surface? The answer is No, for the
simple reason that there is no canonical way of ordering
surfaces. So the infinitesimal areas may be composed in any order
one likes, and each different ordering will give a different
holonomy for the whole surface.  Unless, that is, the holonomy for
infinitesimal surfaces is either trivial (identity) or
Abelian~\cite{Teitelboim:1986ya}.

One way of defining a surface holonomy is to equate it with the
holonomy of a closed loop around the surface, i.e. by taking the
path ordered exponential of a gauge field along the boundary of the
surface element. If the gauge field is non-Abelian, it follows that
surface holonomy defined this way is well defined only if the gauge
field is flat, i.e. has vanishing curvature. The question that
remains is whether it is possible to give an alternative definition
of the surface holonomy such that it can vanish without forcing the
gauge field to be flat. However there is still the problem that two
infinitesimal squares can be composed in two ways, along a common
edge, and at a common vertex. So it seems clear that holonomy for
surfaces cannot be thought of as elements of a group.  It turns out
that categorical Lie groups, or Lie 2-groups, which naturally have
two types of group composition rules, provide an appropriate
description of surface holonomy. In this paper I discuss and relate
different approaches to defining a surface holonomy in terms of Lie
2-groups. 

In \S\ref{defs}, the definition of a Lie 2-Group, or a categorical
Lie group, is given following~\cite{Baez:2002jn}.
Two dimensional parallel transport
will be described by these structures. One-dimensional flux tubes
are transported along two-dimensional surfaces, whose geometric
composition corresponds to composition of elements in a Lie 2-group.

In \S\ref{holonomy}, the construction of surface holonomy is
briefly described following~\cite{Girelli:2003ev}.
Separate group elements, or holonomies, are associated to the face
and edges of an infinitesimal surface element. The edge holonomy
requires a gauge field, a one-form $A$ valued in a Lie algebra,
while the face holonomy requires a two-form $B\,$ valued in another
Lie algebra. The face and edge holonomies are thus elements of two
different groups, and combine according to the composition laws
of the 2-group when the surfaces are composed.  The total holonomy
of the surface vanishes, $F + B = 0\,.$

In \S\ref{twoconn}, a different approach is employed. Instead of
thinking in terms of parallel transporting infinitesimal strings,
fields are transported along strings as well as along paths between
nearby string configurations. In addition to the connection field
which transports along a string configuration, a second one-form
connection $\bar A$ is introduced, rather than a two-form, for
transporting between nearby configurations. Demanding that such
parallel transport be unambiguous leads to an integrability
condition Involving both $A$ and $\bar A\,.$ Allowing $\bar A$ to
take any value subject to this constraint produces an effective
field theory of two-forms.

The paper ends with some discussions in \S\ref{disc} about the
meaning of surface holonomy and related constructions,
given that category theory does not magically create a canonical
ordering for surfaces.

\section{Lie 2-Group}\label{defs}

Parallel transports of fields defined at a point are described by
groups. It is natural to expect that parallel transport of objects
defined on curves or `strings' will require two groups, one for
comparing the values of a field at nearby points of the string, and
another group for comparisons between nearby string configurations.
It turns out that the two groups combine to form a categorical
group, or a Lie 2-group.

A category consists of a set ${\mathcal O}$ of objects and a set
$\{M_{BA}\}$ of morphisms from the object $A$ to the object $B$ for
all $A, B \in {\mathcal O}$, and a composition rule $\circ$ on the
set $\{M_{BA}\}$ such that $\mu_{CB}\circ\mu_{BA} \in M_{CA}\,$ for
all $\mu_{BA}\in M_{BA}\,$ and $\mu_{CB}\in M_{CB}\,,$ and the
following two conditions hold.

\begin{itemize}
\item Composition is associative, i.e.,
\begin{eqnarray}
\mu_{DC}\circ(\mu_{CB} \circ \mu_{BA}) = (\mu_{DC}\circ \mu_{CB})
\circ \mu_{BA} \,.
\end{eqnarray}
for all $\mu_{BA}\in M_{BA}\,, \mu_{CB}\in M_{CB}\,,
\mu_{DC}\in M_{DC}\,. $

\item Identities exist, i.e., for each object $A$ there is a
morphism $\iota_A$ from $A$ to $A$ such that
\begin{eqnarray}
\mu_{BA}\circ \iota_A = \mu_{BA}\,, \qquad \iota_A\circ\mu_{AC} =
\mu_{AC}\,
\end{eqnarray}
for all $\mu_{BA}\in M_{BA}\,, \mu_{AC}\in M_{AC}\,.$
\end{itemize}
Clearly, a category is a generalization of the concept of a group,
rather a monoid, with objects replacing elements and morphisms
replacing maps.

A Lie 2-group is an example of a category in which the set of
objects and the set of morphisms are both Lie groups, and the
composition of morphisms is a homomorphism. A trivial example of a
Lie 2-group is a Lie group $G\,,$ whose elements are now the
objects, so that ${\mathcal O} = G\,,$ and each morphism takes one
element to another\footnote{Often in the context of this example
${\mathcal O}$ is said to consist of the single object $G\,.$ Then
the morphisms are all identity morphisms on $G\,.$},
\begin{eqnarray}
\mu_{21} = g_2 (g_1)^{-1}\,, \qquad \mu_{21}(g_1) = g_2\,.
\end{eqnarray}

It is easy to see that in general the Lie 2-group has two types of
composition rules. Let the objects be elements of $G\,$ and let the
morphisms be elements of some other Lie group $\widetilde G\,.$
Then each morphism is some $\widetilde g(g_2, g_1)$ which takes
$g_1$ to $g_2\in G\,.$ Since this is a category, a morphism which
takes $g_1$ to $g_2$ can be composed with another which takes $g_2$
to $g_3$, and the resulting morphism takes $g_1$ to $g_3,$
\begin{eqnarray}
\widetilde g(g_3, g_2)\circ \widetilde g'(g_2, g_1) = \widetilde
g''(g_3, g_1)\,.
\end{eqnarray}
This composition rule makes no reference to the fact that
${\mathcal O} = G$ is a group.

On the other hand, consider morphisms $\widetilde g(g_4, g_3)$ and
$\widetilde g(g_2, g_1)$ between different elements of $G$.
Now the idea is that the composition $\widetilde g(g_4,
g_3)\cdot\widetilde g(g_2, g_1)$ in $\widetilde G$ should be a
morphism which takes $g_3 g_1$ to $g_4 g_2$, and this morphism
should be a function of $g_1\,, g_2\,, g_3\,, g_4\,.$ This idea can
be easily implemented if $\widetilde G$ is the semi-direct product
of $G$ with some other group $H\,,$ i.e., the morphisms are
elements of $H\rtimes G\,,$ and the action of the morphism on
elements of $G$ is a homomorphism from $H$ to $G\,.$

Recall that $H\rtimes G\,$ consists of groups $G$ and $H\,,$ along
with an action of $G\,$ on $H\,$ given by $\alpha[g](h)\,,$ and the
composition rule is
\begin{eqnarray}
(h, g)\cdot(h', g') = (h\alpha[g](h'), g g')\,.
\label{0104.semi-direct}
\end{eqnarray}
Note that $\alpha$ is a homomorphism from $G$ to
$\mathrm{Aut}(H)\,,$ the group of automorphisms of $H\,,$ and
therefore
\begin{eqnarray}
\alpha[g](\alpha[g'](h)) &=& \alpha[gg'](h)\,,\nonumber \\
\alpha[g](h)\,\alpha[g](h') &=& \alpha[g](hh')\,.
\end{eqnarray}

Then the Lie 2-group thus defined consists of a pair of Lie groups
$G$ and $H\,,$ a homomorphism $\alpha:G\to \mathrm{Aut}(H)\,,$ and
two types of composition rules: $\cdot\,,$ which is the composition
in the semi-direct product $H\rtimes G$ as given in
Eq.~(\ref{0104.semi-direct})\,; and $\circ\,,$ which is the obvious
`composition of morphisms'
\begin{eqnarray}
(h, g)\circ (\bar h, \bar g) = (\bar h h, g)\,.
\label{0104.vertical}
\end{eqnarray}
The order of the objects on the left hand side is a matter of
convention.
Often the homomorphism $t:H\to G\,$ is made explicit so that the
result of applying the morphism $(h, g)$ on $g$ is written as
\begin{eqnarray}
(h, g): g \mapsto t(h)g\,.
\end{eqnarray}
I will usually keep the homomorphism implicit and write $t(h)g$
simply as $hg\,.$ Note also that the composition of morphisms in
Eq.~(\ref{0104.vertical}) is defined only if the two elements $(h,
g)$ and $(\bar h, \bar g)$ are composable, meaning $\bar g = hg\,,$
so that the second morphism can act on the result of the first
one. An important property of a 2-group, which can be checked
directly, is that an exchange law is satisfied,
\begin{eqnarray}
(f_1 \cdot f_2)\circ (f_3 \cdot f_4) = (f_1\circ f_3)\cdot (f_2\circ
f_4)\,,
\label{0104.exchange}
\end{eqnarray}
where $f_1 = (h_1, g_1)\,,$ etc.

Many of the interesting examples of Lie 2-groups are for the case
where the automorphism $\alpha[g](h)$ can be written as $ghg^{-1}\,.$
For convenience of calculation, I will consider only these cases
below, and write the composition of the semi-direct product
accordingly as
\begin{eqnarray}
(h, g)\cdot(h', g') = (h g h' g^{-1}, g g')\,.
\label{0104.sd2}
\end{eqnarray}
%
Finally, a Lie 2-group is equivalent
to what is called a Lie crossed module~\cite{Baez:2002jn}, and in what
follows the two phrases can be used interchangeably.

\section{Surface holonomy}\label{holonomy}

Usual gauge theories are theories of point particles, which 
are described by fields valued in the Lie algebra of $G\,.$ 
The connection one-form or gauge field $A$ parallel transports a 
field along infinitesimal paths. This means that the field changes 
by the action of a group element equal to the path ordered 
exponential of $A$ along a continuous curve. For parallel transport 
of a string the corresponding object should be an element of a Lie
2-group. Such an object can be defined directly~\cite{Baez:2002jn,
Girelli:2003ev, Pfeiffer:2003je}, which I now proceed to describe.

\begin{figure}[htbp]
\includegraphics[width=0.9\textwidth]{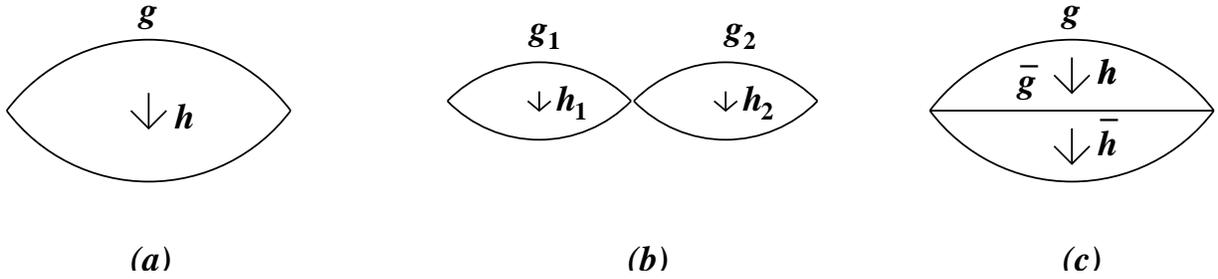}
\vspace*{8pt}
\caption{({\it a}) A bi-gon, ({\it b}) horizontal composition,
({\it c}) vertical composition.  \label{bigon}}
\end{figure}

Let me start with an infinitesimal string.
Consider parallel transporting this string infinitesimally, keeping
the end points fixed. This results in the pair of configurations
schematically drawn in Fig.~\ref{bigon}({\it a}), which will be
called a bi-gon.  This object can be associated with an element
$(h, g)$ of a Lie 2-group by first thinking of a string
configuration in terms of its associated holonomy, i.e. as an
element $g\in G\,.$ Then parallel transporting the string can be
thought of as a morphism, so that an element $h\in H$ needs to be
associated with the surface element bounded by the two
configurations.

For concreteness, let me think of the top edge as the `initial'
configuration, which is then parallel transported (morphed) to the
bottom or final configuration, using an element $h\in H\,.$
Similarly, when constructing holonomies $g\in G$ for the edges, I
will think of the edges as being directed from left to right. The
morphism from the top edge to the bottom edge is a homomorphism
$t:H\to G\,,$ so that its action on the holonomy of the top edge
can be written as $t(h)g\,,$ or as $hg$ as mentioned earlier.  Such
a bi-gon will be termed as `carrying' $(h, g)\,.$ In general, the
holonomy along the bottom edge of a bi-gon need not be the same as
the parallel transported holonomy of the top edge, but two bi-gons
can be composed along a common edge only if these two things are in
fact equal, as explained below.

There are two types of compositions for bi-gons, the horizontal and
vertical compositions, shown in Fig.~\ref{bigon}({\it b}) and
Fig.~\ref{bigon}({\it c}), respectively. For horizontal composition
in Fig.~\ref{bigon}({\it b}), going along the top edges I should
find a composition of the corresponding edge holonomies in $G\,.$
Similarly going along the bottom edges gives a composition of the
morphed holonomies. There should be a corresponding morphism, which
takes the top product to the bottom product, and this should be
made of the two individual morphisms. For the vertical composition,
there is a crucial condition. The holonomy of the bottom edge of
the upper bi-gon in Fig.~\ref{bigon}({\it c}), which results from
morphing the top edge holonomy, must be the same as the top edge
holonomy of the lower bi-gon. That is, two bi-gons carrying $(h,
g)$ and $(\bar h, \bar g)$ can be composed as in
Fig.~\ref{bigon}({\it c}) only if $\bar g = t(h)g\,.$ Otherwise the
composition of the two bi-gons cannot make sense.

It should be now quite obvious how to relate the bi-gons to Lie
2-groups. A bi-gon carrying $(h, g)$ is to be identified with the
element $(h, g)$ of a Lie 2-group, in the same sense a curve can be
identified with its holonomy which lives in some group $G$.
Horizontal composition, as in Fig.~\ref{bigon}({\it b}), is to be
thought of as the product of morphisms given by the composition
rule of the semi-direct product of $H\rtimes G$ as in
Eq.~(\ref{0104.sd2})\,. Vertical composition of bi-gons as
in Fig.~\ref{bigon}({\it c}), whenever composable, is to be thought
of as a composition of morphisms, i.e. as given in
Eq.~(\ref{0104.vertical})\,. The exchange law ensures that
(composable) bi-gons may be composed in any order with the same
result.  Thus the holonomy for a finite surface, between two
configurations of a finite string with the same end points, may be
computed by breaking up the surface into infinitesimal bi-gons and
composing their surface holonomies as elements of a Lie 2-group.

This is an obvious generalization to categories of composing
holonomies along infinitesimal line elements to get the holonomy of
a finite curve. In the latter case the holonomy is the path ordered
exponential of one-form connection or gauge field $A\,,$ valued in
the Lie algebra of some group. The infinitesimal holonomy along a
curve of length $\epsilon$ and tangent $\tau^\mu$ is $1 +
\epsilon\tau^\mu A_\mu\,.$ To generalize this to bi-gons, two
objects are needed, one for the holonomy along an edge, another for
the morphism between edges. So let me introduce a one-form gauge
field $A$ valued in the Lie algebra of $G\,$ in order to compute
the holonomy along an edge. In addition, let me also introduce a
two-form field $B$ valued in the Lie algebra of $H\,.$

Then the holonomy along an infinitesimal edge is of the form $g
\sim 1 + \epsilon\tau^\mu A_\mu \sim 1 + \int A\,$ as before. I
have written an integral because it does not make sense to
represent the edges of bi-gons by tangent vectors. In fact,
calculations become easier if a bi-gon is replaced by a triangle
with the base being identified as the lower edge. Then the
`integral' is the sum of $\epsilon\tau^\mu A_\mu$ terms. Now there
is also a contribution from the infinitesimal surface, of the form
$h\sim 1 + a^2\sigma^{\mu\nu} B_{\mu\nu}\,.$ Here $\sigma^{\mu\nu}$
is the tensor characterizing the surface, and $a^2\sim
O(\epsilon^2)$ is its area. Let me write this holonomy as $1 + \int
B$ in analogy with the one-form. If $g$ belongs to the top edge of
the bi-gon, the morphism to the bottom edge takes $g$ to
\begin{eqnarray}
hg \sim (1 + \int B)\,(1 + \int A) \sim 1 + \int B + \int A\,.
\end{eqnarray}
Since each `integral' is actually an infinitesimal itself, their
product can be ignored to the order of the area. Further, $A$ and
$B$ do not belong to the same Lie algebra, but the homomorphism $t$
induces an obvious map so that $B$ can be brought to the same space
as $A$ and added. Just as curves are identified with group elements
via a gauge field, this completes the identification of bi-gons
with elements of a Lie 2-group, via a pair of `connection' fields
$(A, B)\,.$

\section{Two connections for 2-group}\label{twoconn}

One can take an alternative approach to constructing a surface
holonomy, perhaps somewhat closer in philosophy to quantum field
theories. In this approach, briefly described earlier
in~\cite{Lahiri:2003vm}, one introduces two one-form gauge fields,
valued in the Lie algebras of two groups $G$ and $H\,,$ rather than
a one-form and a two-form.  Then instead of surface holonomy, one
considers the holonomy between identified points on nearby string
configurations. Any surface can be decomposed into a sum of
infinitesimal squares, and thus the result of parallel transporting
a field along an arbitrary path on any surface is unambiguous 
if and only if a certain integrability condition holds. Further, 
it is possible to think of a field theoretic action on which this
condition is imposed as a constraint. The corresponding Lagrange
multiplier field is a two-form in four dimensions (a $(D-2)$-form 
in $D$ dimensions), leading to the usual gauge theories of a
non-Abelian two-form field.  The composition of parallel transports
around squares again follow the structure of a Lie 2-group.

\begin{figure}[htbp]
\includegraphics[width=0.9\textwidth]{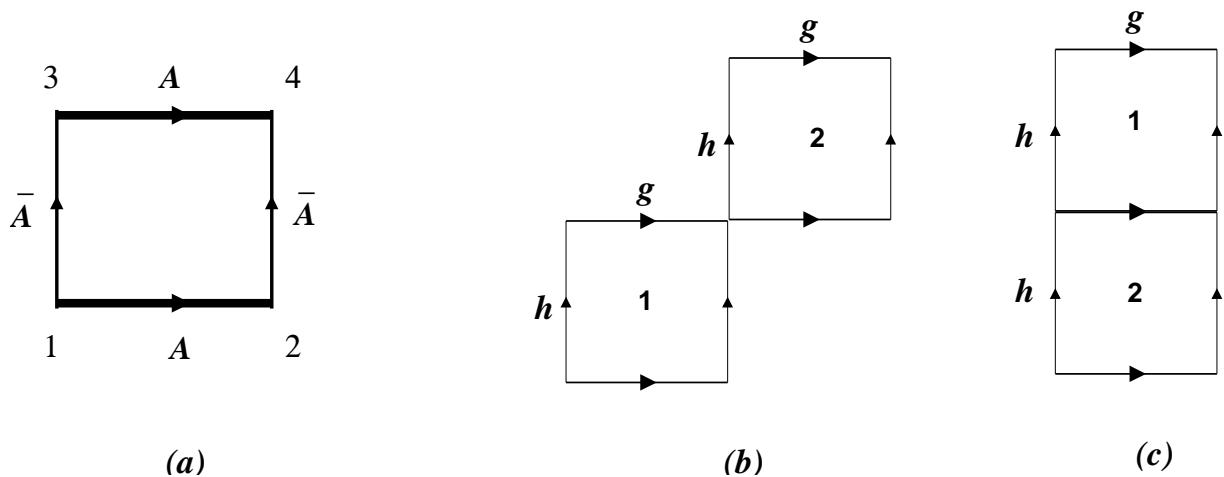}
\vspace*{8pt}
\caption{({\it a}) A square with two connections, ({\it
b}) horizontal composition, ({\it c}) vertical composition.
\label{squares}}
\end{figure}

In this section parallel transport will always mean that of some
field along a curve. Consider an infinitesimal piece of a string,
or flux tube, and another one infinitesimally close to the first
one.  These are the string configurations. The pieces are directed,
and there is a notion of going continuously from one to the
other. This produces the picture of a square, as in
Fig.~\ref{squares}({\it a})\,.
In this square, the top and the bottom edges belong to different
string configurations. For smooth string configurations, it is
possible to unambiguously define vectors along the string and
normal to the string, so squares as these can always be drawn in
such cases.

Consider fields living on the string. Parallel transport is always
integrable in one dimension, so knowledge of the field at any point
on the string determines the field at any other point. Thus given a
connection on the string, any field is well defined at every point
of the string, and can be calculated in terms of its value at some
`zero point' on the string. Now consider an infinitesimally close
string configuration. Again parallel transport along the string
determines the field at any point of the string, in terms of its
value at the transported zero point. But now both the field and the
connection (to be used along the string) must have been parallel
transported to the new configuration from the previous one.  This
parallel transport between configurations could have been done by
any connection, not necessarily the one transporting fields along
the string.

So let $A$ transport fields along the string, and let $\bar A$
transport normal to the string. Also let $A$ and $\bar A$ belong to
the Lie algebras of $G$ and $H\,,$ respectively.
Note that $\bar A$ can live in a subalgebra of the Lie algebra to
which $A$ belongs, but not the other way around, for fields living on
the string must remain in the same algebra as $A$ for all
configurations.

Let me assume for the moment that the two Lie algebras are in fact
isomorphic, so that I can write $\bar A = A + V\,,$ treating the
isomorphism as an equality. Then in Fig.~\ref{squares}({\it a}) a
field can be parallel transported form vertex 1 to vertex 4 either
along the bottom and right edges, or along the left and top
edges. The results of parallel transporting a field using these two
routes around the infinitesimal square will be the same if
\begin{eqnarray}
F_{\mu\nu} + \frac12\left(\partial_{[\mu}V_{\nu]} + \left[A_{[\mu},
V_{\nu]}\right]\right) = 0\,.
\label{0104.integrable}
\end{eqnarray}
If I used the same connection $A$ for all sides of the square, I
would have found a condition of vanishing curvature, $F = 0\,.$
Conditions of this type are referred to a integrability conditions
in the literature~\cite{Alvarez:1997ma}. When a connection
satisfies it, the parallel transport of a field between two points
can be `integrated' along any path connecting the two points,
leading to a unique definition of the field at each point.

The integrability condition of Eq.~(\ref{0104.integrable}) should
be interpreted in a similar fashion.
Any field is completely determined at all points of the string by
parallel transport. If the string is moved to a nearby configuration,
the field can be calculated at every point in the new configuration,
without regard to how intermediate configurations were traversed,
provided the integrability condition holds. Then
transporting fields around squares is unambiguous. And any surface can
be broken up  into infinitesimal squares, and thus a field can be
unambiguously transported along paths on finite surfaces as well.

The composition of the squares now follows the rules of composition
in a Lie 2-group, as is easy to see. The integrability condition
allows me to choose any route around a square, so let me choose one
that is the most convenient for the purpose of comparison with the
bi-gon picture.  Let me bring a field from vertex 4 to vertex 1 by
first dragging it left along the top edge then down along the left
edge. Suppose the top edge has a holonomy $g \in G$ and the left
edge has a holonomy $h \in H\,$ for this route. Then the `total'
holonomy along this route is $hg\,,$ where again I have kept the
homomorphism $t:H\to G\,$ implicit.

Suppose I now compose two squares by joining them at a corner as in
Fig.~\ref{squares}({\it b}). The total holonomy for bringing an
object from the top right corner to the bottom left corner, along the
top and left edges of the squares, is then $h_1 g_1 h_2 g_2\,.$ This
is obviously the same as a square with $g_1 g_2$ on the top edge and
$h_1 g_1 h_2 g_1^{-1}$ on the left edge. Clearly this can be
identified with the product of morphisms as in Eq.~(\ref{0104.sd2}).
On the other hand, if I compose two squares along an edge as in
Fig.~\ref{squares}({\it c}), the holonomy from the top right corner
to the bottom left corner is $h_2 h_1 g_1\,,$ same as that for a
rectangle with $g_1$ on the top edge and $h_2 h_1$ on the left edge.
This can be identified with the composition of morphisms in the Lie
2-group, as in Eq.~(\ref{0104.vertical}). Also quite obviously, these
compositions of squares are exactly the same as the horizontal and
vertical compositions of bi-gons.

The integrability condition is the only one which restricts the
choice of connection $\bar A$ for transporting fields between
strings. Then using a principle typical to quantum theory, I can
sum over all possible choices of the second connection. In other
words, suppose I start with the free action of the gauge field
$A\,.$ When I write the path integral for this action, I should
also integrate over $V\,.$
Then I impose the integrability condition as a constraint on this
path integral to get
\begin{eqnarray}
Z = \int {\cal D}A\, {\cal D}V\, \delta\Big[F + \frac12\,{\rm d}_A
V\Big]\,\exp(-i\int \frac12 F\wedge *F)\,.
\end{eqnarray}
The Lie algebra indices have been summed over as usual. The
$\delta$-functional which enforces the constraint on the theory,
can be rewritten by introducing a Lagrange multiplier field $B\,.$
Then the path integral becomes
\begin{eqnarray}
Z = \int {\cal D}A\, {\cal D}V\, {\cal D}B\,  \exp\left[
-i\int\left(\frac12 F\wedge *F  - B\wedge( F + \frac12\,{\rm d}_A V
)\right)\right]\,,
\end{eqnarray}
It is easy to integrate out $V$ from this path integral, and the
result is a constraint ${\rm d}_A B = 0\,,$ which is imposed on 
a theory with action $I = \int (-\frac12 F\wedge*F + B\wedge F)\,.$

Alternatively I can choose to take a Gaussian average over $V\,.$
This is the same as saying that the second connection $\bar A$ is
peaked around $A\,,$ or that $V$ is peaked around zero. Then the
path integral includes a term proportional to $V^2$ in the
exponent, and can be written as
\begin{eqnarray}
Z = \int  {\cal D}A\, {\cal D}V\, {\cal D}B\,  \exp \left[-i\int
\left(\frac12 F\wedge *F + \frac12 m^2 V^2 - m\,B\wedge( F +
\frac12\,{\rm d}_A V)\right)\right]\,.
\end{eqnarray}
Here $m$ is a constant of mass dimension one, introduced so that
the dimensions of all terms agree, and I have also rescaled $B\to
mB$ so that $B$ has the same dimensionality as the gauge field
$A\,.$ If $V$ is now integrated over, the result is the path
integral
\begin{eqnarray}
Z = \int {\cal D}A\, {\cal D}B\, \exp(i I_{\rm eff})\,,
\end{eqnarray}
with
\begin{eqnarray}
I_{\rm eff} = \int \left(-\frac12 F\wedge *F - \frac12 H\wedge *H + m
B\wedge F\right)\,,
\end{eqnarray}
where $H = {\rm d}_A B\,$ is the field strength of $B\,.$ This
action has several interesting physical consequences, including the
appearance of a pole in the propagator of the gauge field without a
residual Higgs field~\cite{Lahiri:1992hz}.

\section{Discussion}\label{disc}

There is something seemingly very odd about the relation between
surfaces and Lie 2-groups. Recall that the lack of a canonical
ordering for surfaces implied that surface holonomy was not well
defined unless it was trivial or Abelian. Does the bi-gon
construction in terms of Lie 2-groups now allow non-trivial
non-Abelian surface holonomy? There are two ways of answering this
question. Suppose I forget about the category structure of Lie
2-groups and na\"\i vely associate a one-form $A$ to edges and a
two-form $B$ to faces, valued in the Lie algebras of two groups $G$
and $H\,,$ respectively. The holonomy around an infinitesimal
closed loop can be written in terms of the surface it encloses, as
$g\sim 1 + a^2\sigma^{\mu\nu}F_{\mu\nu}\,,$ where $F$ is the
curvature or field strength of $A\,.$ The total surface holonomy is
then $1 + a^2\sigma^{\mu\nu}(F_{\mu\nu} + B_{\mu\nu})\,,$ where as
before $B$ is brought to the same space as $A$ before
addition. This is then the object associated with an infinitesimal
surface, and a product of these objects must be taken when
composing infinitesimal surfaces.  Since there is no canonical
ordering for surfaces, the infinitesimal surfaces may be composed
in any order one likes, and the product of the corresponding
holonomies must give a unique result irrespective of the
order. Clearly, this can happen only if the total holonomy is
trivial or Abelian, i.e. either $F + B = 0$ or the sum lives 
in an Abelian algebra.

But surely the bi-gon construction showed that the ordering of the
infinitesimal surfaces did not matter when composing the elements
of the Lie 2-group? After all, the two-form $B$ was introduced just
for this purpose!  It is true that any surface can be decomposed
into infinitesimal bi-gons --- simply flatten the bottom edge of a
bi-gon, bend the top edge sharply rather than smoothly, and the
bi-gon becomes a triangle, and any surface can be broken up in
triangles.
It is also true that given such a decomposition, the corresponding
surface holonomies will compose as in a Lie 2-group, and the
exchange law of Eq.~(\ref{0104.exchange}) ensures that I can
compose the bi-gons in any order I like, leading to the same final
result. However, there is no contradiction, because not any
arbitrary decomposition is allowed. Only a decomposition in which
adjacent bi-gons are composable, is acceptable. And the `zero
curvature' condition $F + B = 0\,,$ is a direct consequence of the
condition of composability, $\bar g = t(h) g\,,$ where $g$ and
$\bar g$ are the holonomies of the top and bottom edges and $t(h)$
is the contribution from $B$ to the surface
holonomy~\cite{Girelli:2003ev}. 

If the two-form $B$ had not been introduced, the infinitesimal
surface holonomy would be simply $1 + a^2\sigma^{\mu\nu}
F_{\mu\nu}\,,$ and the ordering independence of surface composition
would imply that either $F = 0$ or $A$ is Abelian. Similarly, if I
tried to define a surface holonomy by only a two-form $B$ and
ignored the possibility of composing holonomies along the edges, I
would find that either $B = 0$ or $B$ is Abelian. This result is 
the original one due to Teitelboim~\cite{Teitelboim:1986ya}.

The same sort of argument holds, even more transparently, in the
construction of Lie 2-groups based on two connections. In this case
the important object is the holonomy around surfaces. Any surface
can be decomposed in terms of infinitesimal squares, but the result
of transporting a field along the boundary of the surface is
uniquely defined if and only if the integrability condition holds.
Either way, there is no non-trivial surface holonomy which belongs
to some non-Abelian group.  The real issue is of course whether it
is possible to define a (trivial or Abelian) surface holonomy
involving a non-Abelian gauge field $A$ which is not flat, i.e. for
which $F \neq 0\,.$ This is obviously true for both the procedures
I have considered, and this is what distinguishes the 2-group
construction from the usual results for the integrability of
parallel transport~\cite{Alvarez:1997ma}.

Let me end with a comment about the relationship of surface
holonomy with field theory. For the construction with two
connections, a field theory of non-Abelian two forms appeared
almost naturally by imposing the integrability condition
Eq.~(\ref{0104.integrable}) as a constraint on usual Yang-Mills
theory. For the bi-gon construction, a two-form field is already
present. So it is tempting to try to
derive the flatness condition $F + B = 0\,$ as an equation of
motion in some field theory. Unfortunately the simplest such theory
is rather trivial, with action
\begin{eqnarray}
\int (B\wedge F + \frac12 B\wedge B) \simeq -\int \frac12
F\wedge F\,.
\end{eqnarray}
Of course it is possible to write down other actions using $A$ and
$B\,,$ including the actions found in \S\ref{twoconn}\,. But it is
only in four dimensions that the Lagrange multiplier field $B$ of one
construction has the same structure as the surface gauge connection
$B$ of the other one.

\section*{Acknowledgments}
This paper was written in response to an invitation to the
{\it International Journal of Geometric Methods in Modern Physics},
special issue dedicated to {\em Geometry of Gauge Fields (50 years
of gauge theory)}. I thank the editor, Prof.~G.~Sardanashvily, for
the invitation.





\begin{thebibliography}{0}


\bibitem{Teitelboim:1986ya}
C.~Teitelboim,
Phys.\ Lett.\ B {\bf 167}, 63 (1986).


\bibitem{Baez:2002jn}
J.~C.~Baez,
``Higher Yang-Mills theory,''
arXiv:hep-th/0206130.

\bibitem{Girelli:2003ev}
F.~Girelli and H.~Pfeiffer,
``Higher gauge theory: Differential versus integral formulation,''
arXiv:hep-th/0309173.


\bibitem{Pfeiffer:2003je}
H.~Pfeiffer,
Annals Phys.\  {\bf 308}, 447 (2003)

\bibitem{Lahiri:2003vm}
A.~Lahiri,
``Parallel transport on non-Abelian flux tubes,''
arXiv:hep-th/0312112.


\bibitem{Alvarez:1997ma}
O.~Alvarez, L.~A.~Ferreira and J.~Sanchez Guillen,
Nucl.\ Phys.\ B {\bf 529}, 689 (1998).

\bibitem{Lahiri:1992hz}
A.~Lahiri,
``Generating vector boson masses,''
arxiv:hep-th/9301060.



\end{thebibliography}
\end{document}